\documentclass{articlek}

\renewcommand{\refname}{REFERENCES}
\def\be{\begin{equation}}
\def\ee{\end{equation}}

\def\BibTeX{{\rm B\kern-.05em{\sc i\kern-.025em b}\kern-.08em
            T\kern-.1667em\lower.7ex\hbox{E}\kern-.125emX}}

\usepackage{graphicx}
\begin{document}
\sloppy
\twocolumn[{
{\large\bf DYNAMICS OF ELECTRONS AND AB-INITIO MODELING OF QUANTUM TRANSPORT}\\ 
{\small P. Bokes, peter.bokes@stuba.sk,
Department of Physics, Faculty of Electrical Engineering and Information 
Technology, Slovak University of Technology, Ilkovi\v{c}ova 3, 
812 19 Bratislava, Slovakia}\\


}]
\section{INTRODUCTION}

In the last decade substantial attention has been given to transport
of electrons through individual atomic point contacts or molecules, 
commonly referred to as quantum junctions (QJ).
New experimental results for $I(V)$ curves of QJ are promptly followed 
by their numerical modeling which, however, frequently lead to results 
that differ by order of magnitude from the measured data~\cite{Nitzan03}. 
These discrepancies triggered interest in the reliability of the often 
quietly assumed approximations. Prime suspects are incorrect atomic 
geometries, inappropriate description of the exchange-correlation effects 
based on ground state methods and possible necessity to use a full 
time-dependent simulation to arrive to a correct current-carrying stead-state.

In this short paper we first give a very simple derivation of the 
Landauer formula for a 2-point conductance of QJ $G^{2P}$, based 
on the uncertainty principle. The aim of this is to introduce 
this central equation of quantum transport to a general audience. 
Next we analyse the dynamics of setting up a steady-state current 
in a simple many-electron system and use these observations 
to present physical basis and formal result for the 4-point 
conductance $G^{4P}$, rigorously related to the non-local conductivity 
of an extended system consisting of electrodes and their junction. 

\section{LANDAUER FORMULA AND THE UNCERTAINTY PRINCIPLE}

Let us consider a general QJ. Single-particle quantum states - wavefunctions 
that can be occupied with electrons and carry current - extend with 
nonzero amplitude from the left electrode through the junction into 
the right electrode. These states, as in every metallic system, form 
a continuum of states of energy $E\in (E_{min},E_{max})$, where $E_{min}$ 
and $E_{max}$ are the bottom and the top of the conduction band of 
the electrodes in equilibrium. In equilibrium the Fermi energy $E_{F}$ 
is located somewhere within this interval. Applying bias voltage $\Delta V$ 
between the two electrodes means that those states within the continuum 
that carry current to the right (right-going scattering states) will 
be occupied up to energy $\mu_L$, that is, $e\Delta V$ higher than the states 
that carry current to the left (left-going scattering states), occupied 
up to the energy $\mu_{R}$, i.e. $e\Delta V = \mu_{R} - \mu_{L}$. 
Since degenerate energy levels with both right- and left- going states 
occupied carry zero total current, the only contribution to the transport 
of charge originates from the energy interval $(\mu_{L}, \mu_{R})$, 
occupied only with the right-going states. 

At this point we depart from the traditional derivation~\cite{Buttiker87} 
and make use of the uncertainty principle. Right-going electrons occupying 
states from 
the energy interval $(\mu_{L}, \mu_{R})$ can be put into wavepackets with 
energy uncertainty $\Delta E = \mu_{L} -  \mu_{R} = e\Delta V$. 
The uncertainty principle then states that there is an uncertainty 
in time $\Delta t$ within which we can observe {\it one electron passing 
through the junction}
\be
	\Delta E \Delta t \sim h.
\ee
However, there might be only some probability $T(E_{F})<1$,  
depending in general on the energy\footnote{Assuming smooth dependence of 
the transmission probability on the energy and that the applied bias 
is small we set $T(E)\approx T(E_{F})$ for $E\in(\mu_R,\mu_L)$.}, 
that an electron will pass through 
the junction, out of $N$ electrons approaching the junction within 
time interval $N \Delta t$ only fraction $N T(E_{F})$ will actually 
get through. Using these observation and the fact that current is 
the number of electrons per time, we find 
\be 
	I = 2 e \frac{N \sum_i T_i(E_{F})}{N \Delta t} \sim  \label{eq-landauer}
			\frac{2e^2}{h} \sum_i T_i(E_{F}) \Delta V 
\ee
where we have added a factor $2$ to account for the spin degeneracy and 
sum over possibly several degenerate right-going states $i$. 
This is the celebrated 2 point Landauer formula~\cite{Buttiker87} relating 
the current and the difference between electrochemical potentials in 
electrodes $\Delta V$, the basic equation of modern mesoscopic 
physics. The quantum of conductance $G^0=\frac{2e^2}{h}=(12.9 k\Omega)^{-1}$ 
is obtained for ``open'' QJ with $T_i(E_{F})=1$ for only one $i$. Landauer 
equation represents an important result - it relates quantum-mechanical 
properties of QJ - the transmission probability - with the macroscopically 
measured quantity - the conductance. We would like to note that the weak 
``proportional to'' sign can be made into strong ``equal'' 
introducing occupation-adapted orthogonal wavepacket basis set. 
Including the mean-field, local-neutrality arguments leads
to the 4-point conductance, e.g. in 1D $G^{4P}=G^{2P}/(1-T(E_{F}))$ 
relating the current to the {\it induced electrostatic drop in potential} 
in the vicinity of QJ, $\Delta V^i$~\cite{Buttiker87}.

\section{DYNAMICS OF 1D QUANTUM GAS OF ELECTRONS} 

While previous derivation gives some hint of the time-dynamics in quantum 
transport, namely that based on wavepackets of electrons having significant
amplitude in a given space for certain time $\Delta t$, the approach 
is really describing a steady state. So we ask - how do we make a transition
from equilibrium to steady non-equilibrium situation in QJ?

Consider a 1D electron gas of density $n$, fixed by the Fermi energy
$E_{F} = \frac{\hbar^2 k_F^2}{2m}$. At time $t=0$ we apply  
a localized electric field of the form 
\be 
	E^e = - \frac{\Delta V}{a} (\theta(x+a/2) - \theta(x-a/2) ),
\ee
where $\theta()$ is the unit step function, $a$ the distance on which the field
is nonzero, modeling the width of QJ, and $\Delta V$ corresponds 
to the applied voltage. The response of the density and the current 
can be found using the linear response theory~\cite{Bokes04}. In general, 
the non-local conductivity $\vec{\vec{\sigma}}$ gives a causal linear 
relationship between the total electric field and the current density 
\be
	\vec{j}(\vec{r},t) = \int_{0}^{t} dt' \int d^3 r' 
		\vec{\vec{\sigma}}(\vec{r}, \vec{r}';t-t') \cdot 
		\vec{E}(\vec{r}',t')
\ee 
For our simple system the nonlocal conductivity, calculated directly 
from the continuum of occupied quantum states, is known 
analytically~\cite{Bokes04}. Using this within the general formula (4) 
and performing a simple numerical calculation we find that the current 
is indeed settling to a steady value $I=\frac{2e^2}{h} \Delta V$ 
with a well defined relaxation time $\tau$ (see inset in Fig.1). 
It is instructive to analyze the dependence of the latter on the width 
of the junction $a$, while keeping the overall bias voltage $\Delta V$ 
constant. The resulting dependence
for a gas with $E_{F}=0.07 \textrm{Ha}=2$eV (corresponding to a gold nanowire)
is shown in the Fig.1. The behavior of $\tau$ has a nice 
%
\begin{figure} [h,t]                     
\begin{center}                        
\includegraphics[width=80mm]{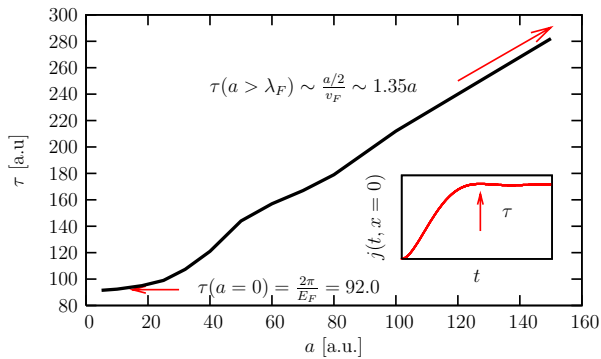}   
\end{center}                         
\vspace{-2mm} \caption{Dependence of the relaxation time $\tau$ (defined by 
the first local maximum in $j(t)$, see the inset) on the width of the 
junction $a$. The arrows show the limiting behaviour.}
\end{figure}                        
physical interpretation: for junctions smaller than the Fermi wavelength 
$\lambda_F$, the relaxation time is constant and given by the 
timescale dictated by the Fermi energy, for junctions larger than 
Fermi wavelength the relaxation is related to a time it takes for 
an electron with Fermi speed to pass the region of the junction.
These limiting results are obtainable by a detailed analysis of the 
analytical formulas as well. 

These results demonstrate the mechanism how the current is established 
for non-interacting electrodes, which is pertinent to all present 
ab-initio calculations. However, as soon as we include even Hartree 
interaction, the process changes dramatically. Namely, the observed current 
in Fig.1 leads to charge imbalance - there are electrons piling up on 
the right of the junction and electrons being depleted on the left of 
the junction. Physically this is almost obvious as any localised 
electric field should be screened out by electrons. The screening 
can be overcome only in the limit $a \rightarrow \infty, E^e \rightarrow 0, 
\Delta V = const$ when the field becomes homogeneous. However, as 
the Fig.~1 clearly shows, in this case the relaxation time becomes 
infinitely as well!  

The resolution is following. To set a steady current in a system 
of interacting electrons, we need to switch on a external homogeneous 
field of {\it finite magnitude for a finite time} $t<t_e$. 
This is to be compared with the infinitesimal magnitude and for all 
positive time in the previous paragraph. In the semi-infinite 
electrode this will create an uniform current even for $t>t_e$ 
and there is no problem with charge accumulation. Only in the region of 
QJ the charge will pile up on the left and deplete on the right.
This will lead to a {\it induced field } $E^i$, localised at the QJ 
and characterised with some induced drop in potential $\Delta V^i$. 
This induced field will self-consistently develop to such a form that 
the current $I$, established throughout the semi-infinite electrodes, 
will be able to pass the junction.

\section{GENERAL EXPRESSION FOR CONDUCTANCE} 
The above ideas can be nicely formalised within the framework of 
the linear response theory. The current is given by
\be 
	I(x,t) = \int_0^t dt' dx' 
		\sigma(x,x';t-t')\left( E^e(x',t') + E^i(x',t') \right).
\ee
Dividing the conductivity into the electrode's, translationally 
invariant part and the part primarily due to the junction, 
$\sigma(x,x') = \sigma^H(x-x') + \sigma^J(x,x')$ respectively,
and performing the long-time analysis we arrive at the final expression
for the steady-state current~\cite{Bokes06}
\be
	I = \frac{\alpha}{\beta} G^{2P} \Delta V^i = G^{4P} \Delta V^i, 
	\label{eq-1}
\ee
where
\begin{eqnarray}
	\mathcal \alpha &=& \sigma^H(q=0,t\rightarrow \infty), \\
        \mathcal \beta  &=& - \int dq \sigma^J(q,q'=0;t\rightarrow \infty), \\
	G^{2P} &=& \int \frac{dq dq'}{2\pi}
			\sigma(q,q';t\rightarrow \infty),
\end{eqnarray}
where $q,q'$ are wavenumbers, reciprocal to $x,x'$ respectively.
The equation (\ref{eq-1}) is expressing the 4-point conductance 
in terms of the limiting character of the non-local microscopic conductivity. 
The utility of this formulation lies in a formally straightforward 
evaluation of the latter using various ab-initio and, unlike the 
Eq.~(\ref{eq-landauer}), also correlated many-body methods.

\noindent ACKNOWLEDGMENT: The author acknowledges Rex Godby, Hector Mera 
and Jeil Jung for stimulating discussions. This research was supported by 
the Slovak grant agency VEGA (project No. 1/2020/05) and the NATO 
Security Through Science Programme (EAP.RIG.981521).


\begin{thebibliography}{99}

\leftskip=-5pt \vspace{-0.3truecm}
\bibitem{Nitzan03}  A.~Nitzan and M.~A.~Ratner, Science {\bf 300}, 1384 (2003).
\bibitem{Buttiker87} M.~B\"{u}ttiker, Y.~Imry, R.~Landauer, and S.~Pinhas, 
	Phys. Rev. B {\bf 31}, 6207 (1985).
\bibitem{Bokes04}  P.~Bokes and R.~W.~Godby, Phys. Rev. B 
		{\bf 69}, 245420 (2004).
\bibitem{Bokes06}  P.~Bokes, J.~Jung, and R.~W.~Godby, cond-mat/0604317 (2006).
\end{thebibliography}
\end{document}